\documentclass[a4paper, times, 10pt,twocolumn]{article}

\usepackage[top=4.9cm,bottom=3.7cm,left=1.5cm,right=1.5cm]{geometry}
\usepackage{ICMLC}
\usepackage{times}
\usepackage{graphicx}
\usepackage{indentfirst}
\usepackage{latexsym}

\usepackage[margin=8pt,font=footnotesize,labelfont=bf,labelsep=period]{caption}

\usepackage{cite}
\usepackage{amsmath,amssymb,amsfonts}
\usepackage{amsthm}
\usepackage{wrapfig}
\usepackage{algorithm}
\usepackage{algpseudocode}

\usepackage{graphicx}
\usepackage{textcomp}
\usepackage{xcolor}
\usepackage{subcaption}
\usepackage{multirow}
\usepackage{booktabs}
\usepackage{soul}
\usepackage{array}
\usepackage{float}
\usepackage{tikz}
\usepackage{longtable}
\usepackage{fancyhdr}
\usepackage{nicematrix}
\usepackage{graphicx}
\usepackage{booktabs}
\usepackage{stfloats}

\DeclareRobustCommand{\hlyell}[1]{{\sethlcolor{yellow}\hl{#1}}}
\DeclareRobustCommand{\hlgreen}[1]{{\sethlcolor{green}\hl{#1}}}
\newcolumntype{P}[1]{>{\centering\arraybackslash}p{#1}}

\begin{document}

\title{A NOVEL APPROACH TO USER AGENT STRING PARSING FOR VULNERABILITY ANALYSIS USING MULTI-HEADED ATTENTION}

\author{\bf \normalsize DHRUV NANDAKUMAR,  SATHVIK MURLI,  ANKUR KHOSLA,  KEVIN CHOI,\\ 
\bf ABDUL RAHMAN, DREW WALSH, SCOTT RIEDE, ERIC DULL,  EDWARD BOWEN \\ 
\\
\normalsize Deloitte \& Touche LLP\\
\normalsize E-MAIL: kevchoi@deloitte.com \\
\\}

\maketitle 
\thispagestyle{empty}

\begin{abstract}
{The increasing reliance on the internet has led to the proliferation of a diverse set of web-browsers and operating systems (OSs) capable of browsing the web. User agent strings (UASs) are a component of web browsing that are transmitted with every Hypertext Transfer Protocol (HTTP) request. They contain information about the client device and software, which is used by web servers for various purposes such as content negotiation and security. However, due to the proliferation of various browsers and devices, parsing UASs is a non-trivial task due to a lack of standardization of UAS formats. Current rules-based approaches are often brittle and can fail when encountering such non-standard formats. In this work, a novel methodology for parsing UASs using Multi-Headed Attention Based transformers is proposed. The proposed methodology exhibits strong performance in parsing a variety of UASs with differing formats. Furthermore, a framework to utilize parsed UASs to estimate the vulnerability scores for large sections of publicly visible IT networks or regions is also discussed. The methodology present here can also be easily extended or deployed for real-time parsing of logs in enterprise settings.}

\end{abstract}
\vspace{5pt}

\begin{keywords}
{User Agent String; Natural Language Processing; Parsing; Transformer}
\end{keywords}

\Section{Introduction}

The increasing reliance on the internet has led to the proliferation of a diverse set of web-browsers and operating systems (OSs) capable of browsing the web. Each of these software packages is versioned separately and has separate security vulnerabilities as an artifact of their design. The vulnerabilities have the potential to be exploited by malicious actors as a method for data theft, network intrusion, or a variety of other tasks. \par

User agent strings (UASs) are a component of web browsing that are transmitted with every Hypertext Transfer Protocol (HTTP) request. They contain information about the client device and software, which is used by web servers for various purposes such as content negotiation and security. UASs contain information pertaining the names and versions of OS, web browser, hardware, and software components used by a client device. This information can be used to assess the vulnerabilities of devices by correlating the extracted information with known Common Vulnerabilities and Exposures (CVEs). These vulnerabilities can then be quantified using the Common Vulnerability Scoring System (CVSS) scores which represent the severity of a vulnerability \cite{cvewebsite}. If done at a larger scale, CVSS scores can be aggregated at regional levels to estimate the vulnerability and security levels of large portions of the public internet. \par

However, the format of UASs is not standardized and can vary greatly, making it difficult to parse and interpret their contents accurately. This can have significant implications for web security, as incorrect information can lead to incorrect decisions being made by web servers, such as serving content that is not compatible with the client device, or failing to properly secure the connection. Furthermore, inaccurate parsing of UASs could significantly impact the estimation of CVSS scores at the device level as CVEs and CVSS scores are heavily reliant on the names and versions of browsers and OSs. \par

Conventional methods for parsing UASs typically rely on rules-based approaches, where regular expressions or heuristics are used to extract relevant security information. While these methods can be effective for well-formed UASs, they are often brittle and can fail when encountering strings with unexpected or non-standard formats. This can lead to incorrect information being extracted or in many cases, not extracted at all. \par

In this paper, we propose a novel approach for parsing UASs using transformer-based natural language processing (NLP) techniques. Our method is designed to be robust and capable of handling a wide range of UAS formats, while accurately extracting information about the client device and software. Furthermore, we also introduce a methodology for estimating the vulnerability of large sections of public internet by correlating UAS information to CVEs. \par

The key contributions of this paper are two-fold:
\begin{itemize}
\item A robust method for parsing UASs using transformer-based NLP techniques, which overcomes the limitations of traditional rules-based methods.
\item The use of extracted fields from the parsing process to assess the vulnerabilities of large portions of the internet by correlating the extracted information with known CVEs, which can help to identify potential security risks and inform mitigation strategies.
\end{itemize}
The accuracy and reliability of UAS parsing is critical for ensuring the security and stability of the internet. By leveraging the information contained in UASs to identify vulnerabilities in networks, our approach has the potential to significantly improve the current state and provide valuable insights into the security of the internet. \par

\Section{Related Work}

UASs have been used extensively to provide content distribution and web servers with the information required for said servers to provide optimal web content. Since the advent of UASs, there has been significant progress in their standardization from a browser perspective. Several browsers and organizations have defined protocols and issued guidance on standardizing UAS formats including measures to shorten UASs and omit non-essential or personally identifiable fields. However, there does not yet exist a universally standardized UAS format. There are several cases when a browser cookie adds its own information in the UAS, or UASs originating from in-app browsers consisting of multiple software names. These cases introduce significant variability to UAS formats which weaken rules-based approaches to UAS parsing; thus, justifying the need for the development of more robust methods. \par

There has been work completed on data relating UASs in similar fashion to the solution proposed in this paper. Crucially however, the scopes of work, methodologies used, and resulting outputs differ from ours.  For example, work has been completed using pre-processing methods related to NLP in the past. Zhange et al. \cite{zhang2015detecting} used context-free grammars to distinguish fake UASs from real ones. Want et al. \cite{wang2017detecting} converted HTTP flow headers into N-gram sequences to obtain a bag-of-words representation. This resulting representation was then fed into a support vector machine to identify malicious network traffic. Tanaka et al. \cite{lightGBMbot} implemented a similar bag-of-words representation, but they used logistic regression, and a tree-based model called Light Gradient-Boosting Machine (LightGBM)\cite{ke2017lightgbm}. \par

UASs have been core source of information where the aim is to detect malicious activities using user logs. The notion of all these works is to extract information from the UAS, encode this information using rule-based techniques and perform statistical analysis on the generated features. For example, Chen et al.\cite{chen2019analysis} used regex on UASs to detect anomalous user agents in network traffic. Boda et al. \cite{boda2012user} described a structure that the majority of UASs they collected follow and leveraged that information for browser fingerprinting. Grill et al. \cite{grill2014malware} divided UASs based on different types, for example legitimate UA, spoofed, empty, etc. Statistical analysis was performed post classification. Lewis et al. \cite{lewis2013http} also uses rule-based approach for parsing, taking into account the defined HTTP structure of UAS. Rule-based approaches have been quite prevalent in this domain. \par

However, multiple works have used NLP-related models on entire HTTP requests to discriminate between malicious and benign traffic. Attempts have been made to perform character level encoding (e.g., 0 for vocabulary characters, 1 for numeric values, etc.) to identify patterns and correlation among multiple UASs. Gao et al. \cite{gao2017anomaly} used this approach along with status code, content length and referrer as feature vectors. Post feature extraction, clustering algorithms were used to identify malicious users. Zolotukhin et al. \cite{zolotukhin2014analysis} used an N-gram model to extract features from UAS, followed by mathematical modeling and Principal Component Analysis (PCA) \cite{jolliffe2002principal} combined with Support Vector Data Distribution (SVDD) \cite{tax2004support}, K-means \cite{ahmed2020k}, Density-based spatial clustering of applications with noise (DBSCAN) \cite{DBSCAN} and aggregated time bin for analysis. Rong et al. \cite{charcnn} used a character-level convolutional neural network, which is a convolutional neural network operating on character-level embeddings of HTTP requests. Park et al. \cite{autoenc} used a similar method, feeding character-level embeddings into an auto-encoder. Gniewkowski et al.\cite{http2vec} leveraged byte-pair tokenization to generate inputs that were then fed into the transformer-based model, a Robustly Optimized BERT Pretraining Approach (RoBERTa) \cite{liu2019roberta}. A key characteristic of these works is they do not extract individual pieces of information from the UASs, but process them as part of a larger feature set. \par

The collective nature of UAS analysis has resulted in reduced focus towards feature extraction from independent UASs. The majority of prior work revolved around identifying patterns from a collection of UASs / web logs. However, in this paper we focus on UAS at rudimentary level to extract features and present threat correlation as an extended use case of this approach. The objectives of our work are to present a foundational approach to UAS parsing and usage in enterprise security and IT settings. \par

\Section{Problem Setup}

In order to build a more robust UAS parser, we propose the use of NLP techniques including Multi-Headed Attention. The scope of our modeling efforts are limited to parsing 4 specific pieces of information from a UAS, although we believe that this methodology can be extended to most other parts of a UAS. Particularly, we focus on extracting:

\begin{itemize}
    \item OS Name and Version
    \item Browser (Software) Name and Version
\end{itemize}

In the following subsections, we explore the structure of our training and validation data, our model architectures, and experimental setup. 

\SubSection{Data and Pre-processing}
Our training dataset consists of over 200 million public UASs collected by WhatIsMyBrowser.com \cite{whatismybrowser_2023}. Each UAS in the dataset was labelled with Software and OS names and versions, along with several other pieces of UAS metadata. We treat this dataset as a labelled dataset with ground truth being the labels assigned by the issuing entity. \par

We pre-processed each UAS to remove special characters, parentheses, remnants of HTML and to standardize whitepsace by performing the substitutions referenced in  Table \ref{tab:CharEdits}. Furthermore, we also limited the length of a UAS to 50 words after preprocessing, wherein any UAS longer than 50 words would be truncated at length 50. \par

\begin{table}[h!]
    \begin{center}
        \caption{Character edits made to format the UASs.}
        \label{tab:CharEdits}
    \vspace{1.5ex}
    \scalebox{0.95}{
    \begin{tabular}{l|l}\hline\hline
       {\bf Original Character}  &  {\bf Replacement} \\\hline
       ``\%20" & `` "\\\hline
       ``\underline{ }" & ``." \\\hline
       ``\textbackslash(" & ``( "\\\hline
       ``\textbackslash)" & ``) "\\\hline
       ``/" & `` " \\\hline
       ``;" & Removed \\\hline
       ``:" & `` : "\\\hline
       ``\%" & `` " \\\hline
    \end{tabular}
    }
    \end{center}
\end{table}

From the pre-processed data, a balanced training dataset was curated for software name and OS name classification. There were a total of 7 classes in each of the Software and OS names. (Software names – Android WebView, Chrome, Facebook App, Internet Explorer, Instagram, Opera, and N/A. OS names – Android, iPad, iOS, Linux, Macintosh, Windows, and N/A). These (top 6) classes were selected based on their popularity among 200M examples. ‘N/A’ class denotes that the UAS belonged to none of the above six classes.  For software name classification and version identification models, each class had 1.4M randomly sampled examples, making the total size of training data almost 9.8M. For OS name classification and version identification models each class had 2M examples and 500k for N/A class, making the total training dataset size as 12.5M. \par

\SubSection{Modeling Considerations and Architectures}

\begin{figure*}[t]
\begin{subfigure}{.53\textwidth}
  \includegraphics[width=.87\linewidth]{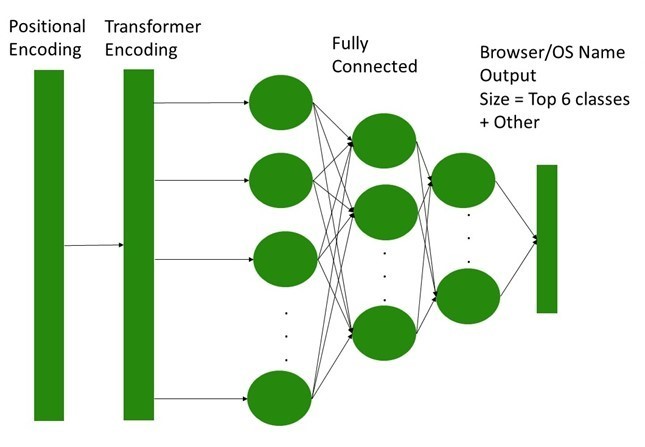}  
  \caption{}
\end{subfigure}
\begin{subfigure}{.53\textwidth}
  \includegraphics[width=.87\linewidth]{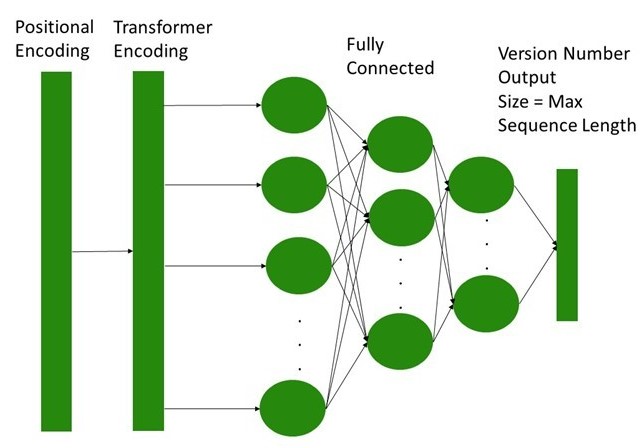}  
  \caption{}
\end{subfigure}
\caption{\centering Model architectures for (a) classification of software name and OS name (b) predicting location of software version and OS  version}
\label{Fig:modelarchs}
\end{figure*}

In order to parse OS and Software names and versions from UASs, we leverage 4 independent models - one for each task. \par

Parsing of OS and Software version numbers is treated as a part of speech (PoS) tagging problem wherein two independent models are trained to output a likelihood of each word in a UAS corresponding to the versions of the OS and Software respectively. Conversely, due to the fact the OS and Software names are often not explicitly stated in UASs, we treat their parsing as a classification problem where two models are trained to classify the names of the target fields using the entire UAS as context respectively. \par
Each of the four models consist of the following three components:

\SubSubSection{UAS Embeddings}
Given that the content of UASs are domain specific, we cannot use pre-existing tokenizers and embedding models to produce vector embeddings of UASs. Furthermore, we also want to leverage information about character substrings in words of a UAS. For example, we want to capture the similarity between the strings \textit{Mac} and \textit{Macintosh} rather than treat them as separate words and tokens. Consequently, we utilize Fasttext \cite{ftext} embeddings trained on a corpus of UASs to represent each word in a UAS as a vector. The Fasttext model is trained using a Continuous Bag Of Words (CBOW) approach and produces a 1-dimensional vector of length 40 per word in a UAS, where a word is any character sequences separated by a space after preprocessing a UAS. Consequently, each UAS will be represented by a vector with length 50 and width 40 after passing through the embedding layer. \par

\SubSubSection{Representation Layer}
Each of the four models share the same architecture at this layer. Embeddings from the UAS Embedding layer are passed through a positional encoding layer followed by a single transformer encoder layer with two attention heads. A positional encoding layer is used in attention-based neural networks to convey the positional information that would not normally be captured by a transformer-based network \cite{vaswani_shazeer_parmar_uszkoreit_jones_gomez_kaiser_polosukhin_2017}.  Our positional encoding layer is defined by the following function: 

\begin{equation}
\label{Eq-1}
\centering{PE_{(pos,2i)} = sin(pos/10000^{2i/d_{model}})}
\end{equation}

\begin{equation}
\label{Eq-2}
\centering{PE_{(pos,2i+1)} = cos(pos/10000^{2i/d_{model}})}
\end{equation}



In which $d_{model}$ refers to the length of the generated word embeddings and $i$ refers to the row upon which the operation is being performed. ${pos}$ refers to the position within each row or sequence that is being passed through the layer. The output of this layer is flattened into a one dimensional vector of length \textit{2000} (from flattening a \textit{50} x \textit{40} vector).

\SubSubSection{Task Specific Heads}
Both the OS and Software name classification heads consist of fully connected dense layers. They are terminated by layers of seven nodes, six of which are for specific classes and the seventh signifying an 'N/A' class. The output of the terminal layers are passed through  a softmax function. The PoS models for version identification also feature fully connected layers but are terminated by a layer with 51 nodes; one for each word in the UAS and another indicating the version was not present in the UAS. The fully connected layers are separated by SeLU activation functions. During the training process, there are also dropout layers that follow the first 3 fully connected layers to combat the risk of overfitting. The overall architecture for each of the models is represented in Figure \ref{Fig:modelarchs}.


\SubSection{Experimental Setup}

Throughout the experiments in this paper, we utilized a 70\%/30\% training and validation split of the data where classes were sampled using Random Stratified Sampling to have appropriate representation in both training and validation data.  We also utilized a fixed batch size of 200 UASs per batch for each of the experiments in this work.

\SubSubSection{Version Indexing models}

The optimizer used in the case of Version Indexing models was Stochastic Gradiatent Descent (SGD) with a learning rate of 0.005 and weight decay of 0.00001. The model was trained using a Cross Entropy loss function. The word at the index with the highest raw value was selected as the version.

\SubSubSection{Name Classification models}

Here, we utilized the SGD optimizer with a learning rate of 0.0005 and weight decay of 0.00001 and a Binary Cross Entropy loss function. A softmax activation function was used in the last layer to compute the probability scores of each class. Class with highest probability was selected as the final output.

\SubSection{Post Processing}

After extraction of the relevant fields from a UAS, our objective is to perform vulnerability analysis on the Classless Inter-Domain Routing (CIDR) ranges from which the UASs originate. We monitor the UASs originating from a particular CIDR range. Once we extract system information (Software and OS) from the UAS, we then estimate a vulnerability score for that particular UAS. For this purpose, we use National Vulnerability Database (NVD: https://nvd.nist.gov/). NVD is a database, maintained by U.S. government, where  Subject Matter Resources (SMRs) analyze the vulnerabilities, based on pre-defined metrics and score them on a scale of 1-10 (10 being highly vulnerable). Although there are mix of old and new scoring systems, such as v2.0 and v3.1, we consider the latest available scores for each vulnerability. \par

These scores can be computed using CVSS. Upon analyzing several factors, such as attack vectors and attack complexity, CVSS computes a score based on base, temporal and environmental metrics (https://nvd.nist.gov/vuln-metrics/cvss/v3-calculator). There are three types of scores provided for each vulnerability – Base Score, Exploitability Score and Impact Score. \par

To map the system information with existing vulnerabilities, we use Common Platform Enumeration (CPE) names (https://nvd.nist.gov/products/cpe). For a single UAS, each of the CPE names and their vulnerability scores are computed using the four-tuple \textit{(OS Name, OS Version, Software Name, Software Version)}. To estimate the vulnerability of a UAS, we compute the mean CVSS score for that UAS accross all known CVEs. This process is conducted for an average Base Score, Impact Score, and Exploitability Score. The averages are given by the equations:

\begin{equation}
\label{Eq-3}
\centering{Vul_{(base\_uas)} = \frac{1}{n} \sum_{i=i}^{n} CVSS_{base_i}}
\end{equation}

\begin{equation}
\label{Eq-4}
\centering{Vul_{(exploit\_uas)} = \frac{1}{n} \sum_{i=i}^{n} CVSS_{exploit_i}}
\end{equation}

\begin{equation}
\label{Eq-5}
\centering{Vul_{(impact\_uas)} = \frac{1}{n} \sum_{i=i}^{n} CVSS_{impact_i}}
\end{equation}


Where \[CVSS_{X}i\] represents the CVSS score for the Base, Exploitability, or Impact score for the \textit{ith} CPE and \textit{n} is the total number of CPEs identified for a UAS.  \par

\begin{figure*}[h]
    \includegraphics[width=\textwidth]{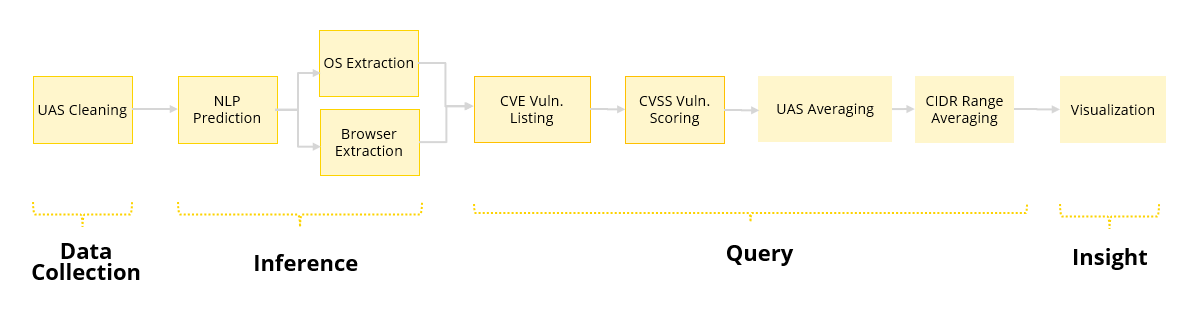}
    \centering
    \caption{Vulnerability Estimation Workflow}
    \label{fig:VulnArch}
\end{figure*}

Once we have the vulnerability scores corresponding to each UAS, we map these scores back to the CIDR ranges. This score distribution is analyzed for each CIDR range to estimate the exposure of the endpoints / systems in that network. We can visualize this vulnerability score distribution against the geographical locations to gauge the exposure on the map. Another way is to analyze this distribution dynamically and visualize how the exposure of a particular CIDR range changes over time. The workflow for the aforementioned vulnerability estimation process is presented in Figure \ref{fig:VulnArch}. NVD Application Programming Interfaces (APIs) were used for CVE Vulnerability Listing and CVSS Vulnerability Scoring. \par

\Section{Results}

Our approach has shown strong results on the 4 tasks we use for evaluation.  For the tasks structured as classification problems, accuracy, precision, recall and f1-scores were generated. For the tasks involving location of version numbers, accuracy is the measure of choice.

\begin{table*}[h]
    \begin{center}
        \caption{ \centering Examples of our model's ability to differentiate between false indicators and the correct answer. A yellow highlight indicates a false indicator, a green highlight indicates the correct answer. The correct answers our model detects are Facebook and Android, respectively.}
        \label{tab:NameClass}
    \vspace{1.5ex}
    \scalebox{0.95}{
    \begin{tabular}{|P{12cm}|P{1.5cm}|P{1.5cm}|P{1.5cm}|P{1.5cm}}\hline\hline
       {\bf UAS}  &  {\bf False Indicator} (Shown in Yellow) & {\bf Correct Indicator} (Shown in Green) & {\bf Generated Answer}\\\hline
       Mozilla 5.0 ( Linux Android 12 M2101K9AG Build SKQ1.210908.001 wv ) AppleWebKit 537.36 ( KHTML, like Gecko ) Version 4.0 \hlyell{Chrome} 104.0.5112.97 Mobile Safari 537.36  \hlgreen{5bFB.IAB} Orca-Android\hlgreen{FBAV} 377.0.0.13.101 5d
 & Chrome & 5bFB.IAB, FBAV & Facebook\\\hline
       Mozilla 5.0 ( \hlyell{Linux} \hlgreen{Andr0id} 10 BRAVIA 4K VH2 ) AppleWebKit 537.36 ( KHTML, like Gecko ) Chrome 84.0.4147.125 Safari 537.36 OPR 46.0.2207.0 OMI 4.21.0.273.DIA6.142
 & Linux & Andr0id & Android\\\hline
    \end{tabular}
    }
    \end{center}
\end{table*}

\SubSection{Name Classification Results}

Tables \ref{tab:SWNRes} and \ref{tab:OPNRes} show the accuracy, precision, and recall of our approach on classification of both OS names and software names. The support column describes the number of rows within the validation set that correspond to each respective class. The results show the model performance is very high on previously unseen data, with the models able to accurately classify strings based on which OS name or browser name they have. Further evaluation runs of the models show their ability to detect strings in which the OS name may be in varying positions or represented as abbreviations or acronyms. Examples of these types of situations are displayed in Table \ref{tab:NameClass}. A yellow highlight indicates a false indicator, a green highlight indicates the correct answer. The correct answers our model classifies are Android and Facebook respectively. The results also show performance does suffer on strings corresponding to Linux OSs, due to its overlap with certain other less frequently encountered OSs that were also UNIX based. Android would be an example of such an OS, but Android OSs were strongly represented in the training data. The results show no such issues with software name classification, which has consistent performance across the 6 most frequently encountered classes. \par

\begin{table}[h]
    \begin{center}
        \caption{Classification of Software Names}
        \label{tab:SWNRes}
    \vspace{1.5ex}
    \scalebox{0.95}{
    \begin{tabular}{l|l|l|l|l}\hline\hline
       {\bf Class Name}  &  {\bf Precision} & {\bf Recall} & {\bf F1 Score} & {\bf Support} \\\hline
       Facebook & 1.00 & 1.00 & 1.00 & 285551\\\hline
       Instagram & 1.00 & 1.00 & 1.00 & 286175 \\\hline
       Chrome & 0.98 & 0.98 & 0.98 & 285615\\\hline
       Android Webview & 0.97 & 0.99 & 0.98 & 283844\\\hline
       Internet Explorer & 0.99 & 1.00 & 1.00 & 285320\\\hline
       Opera & 1.00 & 1.00 & 1.00 & 282903\\\hline
       N/A & 0.98 & 0.95 & 0.97 & 284192\\\hline
    \end{tabular}
    }
    \end{center}
\end{table}
\begin{table}[h]
    \begin{center}
        \caption{Classification of OS Names}
        \label{tab:OPNRes}
    \vspace{1.5ex}
    \scalebox{0.95}{
    \begin{tabular}{l|l|l|l|l}\hline\hline
       {\bf Class Name}  &  {\bf Precision} & {\bf Recall} & {\bf F1 Score} & {\bf Support} \\\hline
       Android & 1.00 & 1.00 & 1.00 & 7038987\\\hline
       iOS & 0.99 & 1.00 & 1.00 & 1281009\\\hline
       Windows & 0.98 & 0.98 & 0.98 & 1006057\\\hline
       Macintosh & 0.96 & 0.90 & 0.93 & 115541\\\hline
       Linux & 0.99 & 0.87 & 0.92 & 116444\\\hline
       iPad & 0.98 & 0.95 & 0.97 & 181760\\\hline
       N/A & 0.87 & 0.91 & 0.89 & 121438\\\hline
    \end{tabular}
    }
    \end{center}
\end{table}

\subsection{Version Indexing Results}

For evaluation of version number identification models, accuracy is our primary measure. Table \ref{tab:INDRes} shows the model performance on both the version number identification tasks. The models have strong accuracy for detecting both kinds of version numbers. Table \ref{tab:INDEx} displays certain OS version numbers that had lower occurrences in the testing set but were still detected by the model. The final row in the table shows the version number with the most occurrences in the testing set for reference. These results display the potential for the models to identify version numbers in varying scenarios and formats. Thus the model shows very strong coverage for different software types and OSs. 

\begin{table}[t]
    \begin{center}
        \caption{Version Number Indexing Results}
        \label{tab:INDRes}
    \vspace{1.5ex}
    \scalebox{0.95}{
    \begin{tabular}{l|l}\hline\hline
       {\bf Task}  &  {\bf Overall Accuracy}  \\\hline
       Software Version Number & 99.4\% \\\hline
       OS Version Number & 99.5\% \\\hline
    \end{tabular}
    }
    \end{center}
\end{table}
\begin{table}[t]
    \begin{center}
        \caption{Version Number Indexing Examples}
        \label{tab:INDEx}
    \vspace{1.5ex}
    \scalebox{0.95}{
    \begin{tabular}{l|l|l|l|l}\hline\hline
       {\bf Version Number}  &  {\bf Precision}  &  {\bf Recall}  &  {\bf F1 Score}  &  {\bf Support}\\\hline
       2022060972 & 0.996 & 0.99 & 0.993 & 1732 \\\hline
       21.113 & 0.948 & 0.998 & 0.973 & 2341 \\\hline
       12.4.6 & 0.998 & 1.000 & 0.999 & 3215 \\\hline
       10.0 & 1.000 & 1.000 & 1.000 & 424575 \\\hline
    \end{tabular}
    }
    \end{center}
\end{table}

\SubSection{Vulnerability Visualization}

Table \ref{tab:cidr_score} shows the Vulnerability Scoring system for a few examples in our data. Six UASs are mentioned for two CIDR ranges. Software name and OS information is extracted from these UASs, and a base vulnerability score is computed for each UAS accordingly. The \textit{Avg. Base Score} column represents average of all the base scores corresponding to a particular CIDR range. Furthermore, the CIDR range to base score mapping can be extended to identification of vulnerable systems in a network. \par

\begin{figure}[t]
\begin{center}
\scalebox{0.35}[0.35]
{\includegraphics{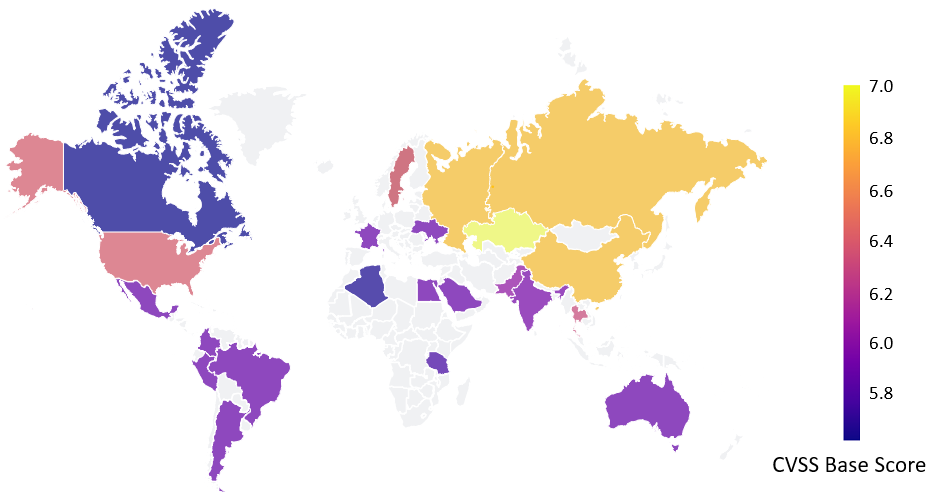}}
\caption{\centering Vulnerability Assessment (on Base Scores) vs CIDR ranges - Geographical analysis. }
\label{Fig:globalanalysis}
\end{center}
\end{figure}

Figure \ref{Fig:globalanalysis} depicts the visualization built on the top of CIDR ranges against Base Vulnerability Scores after post-processing. This visualization is based only on a subset of our data. However, we believe that the illustration is representative of the promise our proposed approach holds for vulnerability monitoring and trend analysis on a per-network or region level. These visualizations  could help security experts analyse changing trends to the security posture of their internal networks or networks to which they connect. It can also assist in the cases where SMRs are attempting to patch vulnerable endpoints in a network.

\begin{center}
\begin{table*}[!t]
 \centering
 \small
 \caption{Scoring system examples on CIDR Ranges}
 \label{tab:cidr_score}
{\footnotesize %
 \begin{NiceTabular}{p{1.7cm}|p{7cm}|p{1.5cm}|p{1.2cm}|p{1.5cm}|p{1.5cm}}[hvlines, left-margin=1em]

	{\bf CIDR Range}  &  {\bf User Agent Strings}  &  {\bf Software}  &  {\bf Operating System}  &  {\bf Base Score (from CPE names)}  &  {\bf Avg. Base Score}\\

	\Block{5-1}{1.123.**.*/24} & Mozilla 5.0 ( Windows Phone 8.1 ARM Trident 8.0 Touch rv : 11.0 IEMobile 11.0 ) ...... & Internet \hspace{1cm} Explorer, 11 & Windows 8.1 & 6.15814104 & \Block{5-1}{5.977965983} \\
	& Mozilla 5.0 ( Linux Android 12 SM-G986B ) AppleWebKit 537.36 ( KHTML, like Gecko ) Chrome 105.0.0.0 Mobile ...... & Chrome, 105.0.0.0 & Android, 12 & 6.57974525 & \\
	& Mozilla 5.0 ( Linux Android 11 SM-G988B ) AppleWebKit 537.36 ( KHTML, like Gecko ) Chrome 105.0.0.0 Mobile ...... & Chrome, 105.0.0.0 & Android, 11 & 6.07651594 & \\
	& ...... & ...... & ...... & ...... & \\
    & ...... & ...... & ...... & ...... & \\

	\Block{5-1}{101.127.**.*/24} & Mozilla 5.0 ( Linux Android 9 SM-N960F Build PPR1.180610.011 wv ) AppleWebKit 537.36 ( KHTML, like Gecko ) Version 4.0 Chrome 105.0.5195.136 Mobile Safari 537.36 SheinApp( shein 8.5.6 ) TTID hybrid@wing.android.1.0.1 ...... & Android WebView, 105.0.5195.136 & Android & 5.79486228 & \Block{5-1}{5.947666269} \\
	& Mozilla 5.0 ( Linux Android 12 SM-A515F Build SP1A.210812.016 wv ) AppleWebKit 537.36 ( KHTML, like Gecko ) Version 4.0 Chrome 105.0.5195.136 Mobile Safari 537.36  5bFB.IAB Orca-AndroidFBAV 379.1.0.23.114 5d ...... & Facebook App, 379.1.0.23.114 & Android, 12 & 5.80187262 & \\
	& Mozilla 5.0 ( Windows NT   10.0 WOW64 Trident 7.0 NMTE rv : 11.0 )...... & Internet \hspace{1cm} Explorer, 11 & Windows, 10 & 6.39906788 & \\
	& ...... & ...... & ...... & ...... & \\
    & ...... & ...... & ...... & ...... & \\

 \end{NiceTabular}
 }
\end{table*}
\end{center}

\Section{Conclusion and Future Work}
In this work, we have introduced a novel, foundational approach to UAS parsing that is scalable, extensible, and most importantly, robust to the varying formats of UASs. With strong performance in several categories, we demonstrate that our proposed approach also shows promise in enterprise settings where reliable parsing of UASs is of paramount importance. \par

We have also demonstrated how our approach could ostensibly be used in cybersecurity-specific contexts for vulnerability trends analysis on a network or regional level. While not in the scope of this work, we believe that our methodology can be adapted and utilized in several other fields of application where the independent data consist of strings that have a complex, partially-standardized structure and cannot be tokenized using tokenizers for common natural languages. \par

As for future work, we believe that there can be several avenues of research for improving the performance of our models. The primary examples of one such direction is improving the performance of our parsers on very-low-support classes, or few-shot learning, for parsing rare UASs such as UASs from smart speakers. We are also looking to improve the performance on models on difficult examples, as seen with the \textit{Linux} OS in our experimental results. We have seen promising initial results along this line of inquiry by utilizing curriculum learning to train our models to parse easier examples first before introducing them to harder examples.  Our approach uses a weighted average of the Euclidean distance of a chosen UAS's word embedding from the mean word embeddings,  the difference in length of the string from the mean length, and the output of the name classification models. Given the initially strong experimental results, we believe that this research direction also warrants further investigation. \par

\bibliographystyle{IEEEtran}


\end{document}